\documentclass{PoS}
\title{Fisher's zeros as boundary of RG flows in complex coupling space}

\ShortTitle{Fisher's zeros and RG flows in complex coupling space}

\author{A. Bazavov $^b$, A. Denbleyker $^a$, Daping Du $^a$, Yuzhi Liu $^a$ \speaker{Y. Meurice } $^a$,  and Haiyuan Zou $^a$\\
 
$^a$Department of Physics and Astronomy, The University of Iowa\\
Iowa City, Iowa 52242, USA\\
E-mail: \email{alan-denbleyker@uiowa.edu}\\
        E-mail: \email{daping-du@uiowa.edu}\\
         E-mail: \email{yuzhi-liu@uiowa.edu}\\
         E-mail: \email{yannick-meurice@uiowa.edu}\\
        E-mail: \email{haiyuan-zou@uiowa.edu}\\
        \\
$^b$Physics Department, Brookhaven National Laboratory\\Upton NY 11973, USA\\
E-mail: \email{obazavov@bnl.gov}
}

\abstract{We discuss the possibility of extending the RG flows to complex coupling spaces. We argue that the Fisher's  zeros
are located at the boundary of the complex basin of attraction of IR fixed points.  We support this picture with numerical calculations at finite volume for $2D$ $O(N)$ models in the large-$N$ limit and the hierarchical Ising model 
using the two-lattice matching method.  
We present numerical evidence supporting the idea that, as the volume increases,  the Fisher's zeros of 4-dimensional pure gauge $SU(2)$ lattice gauge theory with a Wilson action, stabilize at a distance larger than 0.1 from the real axis in the complex $\beta=4/g^2$ plane. We show that when a positive adjoint term is added,  the zeros 
get closer to the real axis. We compare the situation with the $U(1)$ case. 
We discuss the implications of this new framework for proofs of confinement and searches for nontrivial IR fixed points in models beyond the standard model.}

\FullConference{The XXVIII International Symposium on Lattice Field Theory, Lattice2010\\
		June 14-19, 2010\\
		Villasimius, Italy}

\begin{document}

\section{Introduction} 
\begin{figure}[b]
\begin{center}
\includegraphics[width=2.2in,angle=0]{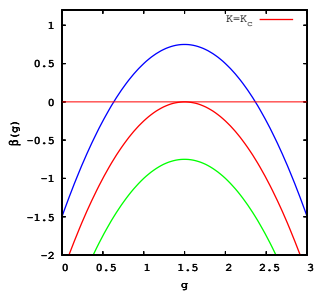}
\caption{ \label{fig:quad}Sketch of what happens when we  reduce  the constant term in a quadratic $\beta$ function: the IR and UV fixed points merge and disappear in the complex plane
}
\end{center}
\end{figure}

There has been a renewed interest in the Lattice community 
\cite{shamir08, appelquist09, hasenfratz09,fodor09, deuzeman09,Myers:2009df,DelDebbio:2010ze}
for the possibility \cite{banks81} of having nontrivial infrared  fixed points in asymptotically 
free gauge theories (see recent reviews by deGrand \cite{DeGrand:2010ba} and Ogilvie \cite{Ogilvie:2010vx}). 
One particularly interesting situation from a phenomelogical point of view is when the $\beta$ function approaches zero from below and encounters only small changes over a significant range of scale. We then say that the ``running'' coupling constant ``walks''. From this perspective, the ability to control the height of the $\beta$ function 
appears to be quite important. A simple model where this can be done easily is the case of  a quadratic $\beta$ function which appears naturally in  the quantum mechanical $1/r^2$ potential and other problems where conformality can be lost and restored
\cite{Kaplan:2009kr,moroz09}.  For a sufficiently large value of the constant term of a quadratic $\beta$ function, 
one  infrared (IR) and one ultraviolet (UV) fixed points are present. By lowering this constant term, we can make the two fixed points coalesce and then disappear in the complex plane. The situation is sketched in Fig. \ref{fig:quad}. 

This motivated us 
\cite{Denbleyker:2010sv} to study extensions of renormalization group (RG) flows in the complex coupling plane. A general feature that we observed is that 
the Fisher's zeros - the zeros of the partition function in the complex coupling plane - apparently act as  ``gates'' for the RG flows ending 
at the strongly coupled fixed point. This can be seen as a complex extension of the general picture proposed by Tomboulis \cite{Tomboulis:2009zz} to prove confinement: the 
gate stays open as the volume increases and the flows starting in a complex neighborhood  the 
UV fixed point can reach the IR fixed point where confinement and the 
existence of a mass gap are clearly present. In general, losing conformality  corresponds to the generation of a mass gap and the presence of confinement and complex fixed points not on the real axis.  We argue that such fixed points are related to the absence of Fisher's zeros on the real axis. 

In the following, we illustrate this scenario 
with model calculations 
for the 
 $2D\ O(N)$ non-linear sigma models in the large-$N$ limit
and the 
Ising hierarchical model.
In all cases, the  Fisher's zeros (of the partition function) seem to govern the global behavior of the flows near the real axis. In the infinite volume limit, these zeros delimit the boundary of the basin of attraction of the strongly coupled fixed point. 
For confining models, a ``gate'' remains open. 
We considered  modifications or deformations that may affect that behavior (finite volume, change of dimension, additional pieces in the action).
We will then present recent results regarding the Fisher zeros for 
 $U(1)$ and $SU(2)$ $4D$ LGT (zeros at different volume; no RG flows yet).
Note that at finite volume all the models  considered here have a partition function analytical in the entire complex $\beta \propto 1/g^2$ plane. 
We should also mention
previous studies of RG transformations  in the complex temperature or coupling plane for 1D spin models and 2D gauge models  \cite{Damgaard:1993df}.
 
\section{Complex RG flows in spin models}
We now discuss numerical calculations of complex RG flows in spin models. We first consider 
the $2D\  O(N)$ non-linear sigma model in the large-$N$ limit. The partition function reads: 
 \begin{equation}
  Z=\int \prod _{\bf x} d^N\phi_{\bf x}\delta(\vec{\phi}_{\bf x}.\vec{\bf \phi}_{\bf x}-1){\rm e}^{-(1/g_0^2)\sum_{{\bf x},{\bf e}}(1-\vec{\bf \phi}_{\bf x}.\vec{\bf \phi }_{\bf x+e})} \ . 
  \end{equation}
We use the notation 
$\beta\equiv 1/(g_0^2N)$, not to be confused with the $\beta$ function, 
for the inverse 't Hooft coupling and $M\equiv m_{gap}/\Lambda_{UV}$ for the mass gap in cutoff units. For 
 large-$N$, we have the gap equation: 
 \def\mn{\mathcal{N}}
 \begin{equation}
\label{eq:gap}
\beta(M^2)=
\int_{-\pi}^{\pi}\int_{-\pi}^{\pi}\frac{d^2k}{(2\pi)^2}\frac{1}{2(2-{\rm cos}(k_1)-{\rm cos}(k_2))+M^2}\  .
\end{equation}
At small coupling we have the asymptotically free relation: 
$\beta(M^2)\simeq 1/(4\pi)\ {\rm ln}(1/M^2)$. 
\begin{figure}[h]
\begin{center}
\label{fig:infon}
\includegraphics[width=2.6in,angle=0]{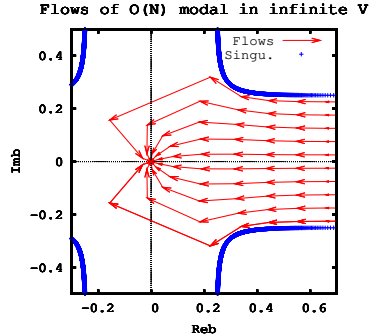}
\caption{ RG flows obtained by rescaling complex values of $M^2$ in Eq. (2.2) . }
\end{center}
\end{figure}

We first consider a simple complex extension of the running of the coupling when the UV cutoff is decreased:  we take the mass gap on a small circle in the complex plane: $m_{gap}=\epsilon {\rm e}^{i\theta}$ and then lower the cutoff. 
$\Lambda_{UV}\rightarrow \Lambda_{UV}/b$. The flows  for $b=2$ are represented by red arrows  in Fig. \ref{fig:infon} . The blending blue crosses (called ``blue lines'' hereafter) are the $\beta$ images obtained from Eq. (\ref{eq:gap}) of two lines of points located very close above and below the $[-8,0]$ cut of $\beta(M^2)$ in the $M^2$ plane. The RG flows are constrained to stay inside the blue lines and we will explain why the Fisher's zeros are expected to stay outside the blue lines. The flatness of the flows at large $\beta$ follows from the asymptotic freedom running: the complex phase in $M^2$ is not affected by the cutoff and its logarithm has a constant imaginary part.  

As explained in Ref. \cite{Meurice:2009bq}, the zeros of the partition function should only appear outside the blue lines in Fig. \ref{fig:infon}. This can be inferred from the representation 
\begin{equation}
\oint_C  d\beta (dZ/d\beta)/Z =i2\pi \sum_q n_q(C)\ ,
\end{equation}
where $ n_q(C)$ is the number of zeros of order $q$ inside $C$. 
For large $N$, 
\begin{equation}
\oint_C  d\beta (dZ/d\beta)/Z \propto\oint_{C'}  dM^2 (d\beta/dM^2) (M^2-1/\beta) 
\end{equation}
The second term has a pole at $\beta=0$, but it is compensated by a pole 
in $M^2$ because for small $\beta$ (large $M^2$), we have $\beta \simeq 1/M^2$ (see (2.2)). At infinite volume, the poles of $(d\beta/dM^2)$ are in the cut (the real interval $[-8,0]$).  If the contour $C'$ in the $M^2$ plane does 
not cross the cut, then there are no zeros of the partition function inside the corresponding $C$ in the $b$-plane.  We conclude that in the large-$N$ limit, there are no Fisher's zero in the  image of the cut $M^2$ plane. Fig. \ref{fig:infon} illustrates a situation that we expect to be generic: the RG flows admit an analytical continuation in the complex plane until Fisher's zeros appear. 

\begin{figure}[b]
\begin{center}
\includegraphics[width=2.2in,angle=270]{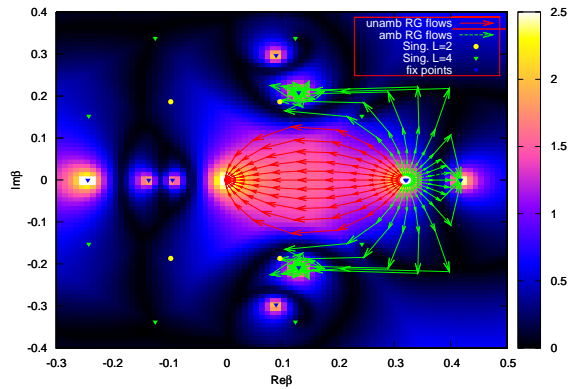}
\caption{\label{fig:oncontour}Complex RG flows with the 2-lattice matching method  for $L=4$; Color scale :-Ln(ambiguity)}
\end{center}
\end{figure}

We also considered the two lattice matching \cite{Hasenfratz:1984hx}. In short, we considered the sums of the spins in four $L/2\times L/2$ blocks $B$. $NB$ denotes a nearest neighbor block of $B$. We define:
\begin{equation}
R(\beta,L)\equiv\frac{\left\langle (\sum_{x\in B}\vec{\phi}_x )(\sum_{y\in NB}\vec{\phi}_y)\right\rangle_\beta}{\left\langle (\sum_{x\in B}\vec{\phi}_x)(\sum_{y\in B}\vec{\phi}_y ))\right\rangle_\beta} \ .
\end{equation}
A discrete RG transformation mapping $\beta$ into $\beta'$ while the lattice spacing changes from $a$ to $2a$  is obtained by   matching: $R(\beta,L)=R(\beta',L/2)$. This method bypasses the difficult construction of the effective hamiltonian and the calculation of the field rescaling. The solutions were found numerically by
Newton's method. When several solutions could be found, we picked the closest to the original $\beta$. We defined the ambiguity as $|\beta-\beta_{closest}|/|\beta-\beta_{2d. closest}|$.  The results are presented in Fig. \ref{fig:oncontour} for $L=4$. 

It should noted that the appearance of a fixed point on the real axis near $\beta\simeq 0.3$ is a finite size effect. As $L$ increases, the fixed point moves to arbitrarily large $\beta$. In general, finite size effects in the complex plane are very intricate since the mapping of Eq. (\ref{eq:gap}) becomes 
a ratio of polynomial and requires a multi-sheet Riemann surface to be inverted. This will be discussed at length in Ref. \cite{onprogress}. 

We also applied the two-lattice matching for  Dyson' s hierarchical model with a Ising measure. 
It is a lattice model with block interactions depending on the details of the block configurations in a minimal way.
In this model, the local potential approximation is exact. 
Its recursion formula is related to Wilson's approximate recursion formula (that allowed the first numerical RG calculations) 
but the exponents are different. 
The probability distribution for the total spin in blocks can be calculated iteratively with this formula.  
The model has a continuous parameter that can be tuned in order to reproduce the scaling of a $D$-dimensional massless Gaussian field. 
For $D=$ 2 Dyson has proven rigorously the absence of transition \cite{dyson69}. For $D=3$, it has a Wilson fixed point near $\beta\simeq 1.179$. 
These facts are reviewed in Ref. \cite{hmreview} where the question of the improvement of the hierarchical approximation is also discussed. 
Complex RG flows showing bifurcations around lines of Fisher's zeros are shown in Fig. 2 of Ref. \cite{Denbleyker:2010sv}. The finite size effects and the ambiguity in the search will be discussed in Ref. \cite{hmprogress}. 

\section{Fisher's zeros in $4D$ LGT}

\begin{figure}
\begin{center}\includegraphics[width=3in]{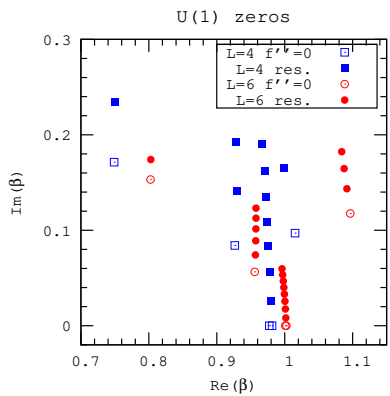}
\caption{\label{fig:abz}Images of the zeros of $f''(s)$ in the $\beta$ plane (open symbols) and Fisher's zeros (filled symbols) for $U(1)$ on $4^4$ (squares) and $6^4$ (circles) lattices.}
\end{center}
\end{figure}

At this point, we have not constructed complex RG flows for gauge theories, but we have designed numerically stable methods to calculate the Fisher's zeros in $U(1)$ and $SU(2)$ $4D$ lattice gauge theories.  We use the 
spectral decomposition
\begin{equation}
Z =\int_0^{S_{max}}dSn(S){\rm e}^{-\beta S}
\end{equation}
with $n(S)$ the density of states and  $\mn$ the number of plaquettes.  For large $\mn$, we have 
\begin{equation}
n(S){\rm e}^{-\beta\mn s}={\rm e}^{\mn (f(s)-\beta s)} ={\rm e}^{\mn (f(s_0)+(1/2)f''(s_0)(s-s_0)^2+\dots)}
\end{equation}
with $s=S/\mn$ and $f'(s_0)=\beta$. $f(s)$ can be interpreted as a color entropy density. 
If $Ref''(s_0)<0$, the distribution becomes Gaussian in the infinite volume limit.  Since Gaussian distributions have no complex zeros \cite{alves91}, the level curve $Ref''(s_0)=0$ is the boundary of the region where Fisher's zeros may appear. They typically delimit thin elongated regions ending at a complex zero of $f''$. Results for $SU(2)$ are shown in Fig. 3 of Ref. \cite{Denbleyker:2010sv}  and can be compared with the $U(1)$ case in  Fig. \ref{fig:abz}. 
The main difference is that the imaginary part of the lowest zero appear to stabilize at a finite, non-zero value for $SU(2)$ while this quantity goes to zero in the $U(1)$ case. The details of the numerical construction of $f(s)$ will be discussed in Ref. \cite{su2progress}. 

In the $U(1)$ case,  multicanonical methods were used \cite{Bazavov:2009pj} and naive histogram reweighting works well. The numerical error $\delta Z$ can be estimated from $(n_i(S)\ - \ <n(S)>)$, where $i$ is an index for independent runs. Zeros can be excluded if $|\delta Z|<<|Z|$.
For $SU(2)$, the imaginary part of Fisher's zeros are too large to use simple reweighting methods. By using Chebyshev interpolation for $f(s)$ and monitoring the numerical stability of the integrals with the residue theorem as in Eq. (2.3), it is possible to obtain reasonably stable results.   Unlike the $U(1)$ case,  the imaginary part of the lowest zeros does not decrease as the volume increases, but their linear density increases at a rate compatible with $L^{-4}$. The effect of an adjoint term (+0.5) is that the lowest zero goes down by about 40 percent. 

\section{Conclusions}
In summary, we have shown that it is possible to extend various RG flows to the complex $\beta$ plane.
When the size of the system is comparable to the Compton wavelength of the gap, different methods can give very different answers. 
In all cases considered, the 
Fisher's zeros control the global behavior of the RG flows. 
Confinement occurs when the zeros leave an ``open gate'' on the real axis. We 
plan to study related questions for QED and $SU(3)$ with various $N_f$.

\begin{acknowledgments}
This 
research was supported in part  by the Department of Energy
under Contract No. FG02-91ER40664. Part of this work was done while Y. M. was at the workshop ``Critical Behavior of Lattice Models'' at the Aspen Center for Physics in May-June 2010. We thank the participants for valuable discussions and suggestions. 
\end{acknowledgments}

\end{document}